\documentclass[twocolumn,aps,pra,showpacs,floatfix]{revtex4}

\usepackage{amsmath}
\usepackage{amsfonts}
\usepackage{amssymb}
\usepackage{amsthm}
\usepackage{color}
\usepackage{dcolumn}
\usepackage{graphicx}
\usepackage{epsf}
\usepackage{amsmath}
\usepackage{bbm}
\usepackage{bm}
\usepackage{times}
\usepackage{nicefrac}
\usepackage{dcolumn}
\usepackage{psfrag}
\newcolumntype{.}{D{x}{}{-1}}

\def\buildrel#1\under#2{\mathrel{\mathop{\kern0pt #2}\limits_{#1}}}

\def\bm#1{\mbox{\boldmath$#1$}}

\def\corresponds{{\lower.2ex\hbox{=}}{\rm\kern-.75em^\triangle}}
\def\succsim{\succ\kern-.9em_\sim\kern.3em}
\def\precsim{\prec\kern-1em_\sim\kern.3em}
\def\slantfrac#1#2{\kern1em^{#1}\kern-.3em/\kern-.1em_{#2}}
\def\lfrac#1#2{{}^{#1\!}\kern-.0em/_{#2}}

\begin{document}

\sloppy

\title{Radiative Corrections to Multi--Level Mollow--Type 
Spectra\footnote{Dedicated to Prof.~H.~Walther on the occasion 
of his 70$^{\rm th}$ birthday.}}

\author{Ulrich D.~Jentschura}
\email{jentschura@mpi-hd.mpg.de}

\author{J\"{o}rg Evers}
\email{evers@mpi-hd.mpg.de}

\author{Christoph H.~Keitel}
\email{keitel@mpi-hd.mpg.de} 

\affiliation{Max--Planck--Institut f\"ur Kernphysik,
Saupfercheckweg 1, 69117 Heidelberg, Germany}

\begin{abstract}
This paper is concerned with two rather basic 
phenomena: the incoherent fluorescence spectrum
of an atom driven by an intense laser field 
and the coupling of the atom to the (empty) 
modes of the radiation field.
The sum of the many-photon processes gives rise
to the inelastic part of the atomic fluorescence,
which, for a two-level system, has a well-known characteristic 
three-peak structure 
known as the Mollow spectrum.
From a theoretical point of view, the 
Mollow spectrum finds a natural interpretation in terms 
of transitions among laser-dressed states which are 
the energy eigenstates of a second-quantized 
two-level system strongly coupled to 
a driving laser field. As recently shown, 
the quasi-energies of the {\em laser-dressed} states
receive radiative corrections which are nontrivially
different from the results which one would expect 
from an investigation of
the coupling of the {\em bare} states to the 
vacuum modes. In this article, we briefly review 
the basic elements required for the analysis of the 
dynamic radiative corrections, and we generalize the treatment of the 
radiative corrections to the incoherent part of 
the steady-state fluorescence to a three-level system
consisting of $1S$, $3P$ and $2S$ states.
\end{abstract}


\maketitle

%
%
\section{\label{intro}Introduction}

The theory of the 
interactions of atoms with light began in the 1920s and 1930s
with the description of a number of basic processes;
one of these is the 
Kramers--Heisenberg formula~\cite{KrHe1925}
which describes a process in which an electron absorbs and emits 
one photon. The corresponding Feynman diagram is shown in
Fig.~1 {\sl (a)}. This scattering process is elastic, the electron 
radiates at exactly the driving frequency, a point 
which has been stressed a long time ago~\cite{We1931}.
If more than one photon is absorbed or emitted,
then the energy conservation applied only to the 
sum of the frequencies of the absorbed and emitted photons
[see Fig.~1 {\sl (b)}]. The frequencies of the atomic fluorescence
photons (of the scattered radiation) are 
not necessarily equal to the laser frequency $\omega_{\rm L}$.
From the point of view of the $S$-matrix formalism, 
Fig.~1 {\sl (a)} and {\sl (b)} correspond to the forward
scattering of an electron in a (weak) laser field.

Indeed, the entire formalism used for the 
evaluation of quantum electrodynamic shifts of 
atomic energy levels is based on the (adiabatically
damped) $S$-matrix theory. The 
Gell-Mann--Low Theorem~\cite{GMLo1951,Su1957} yields the 
formulas for the energy shifts.

%
%
\begin{figure}[t]
\begin{center}
\includegraphics[height=2.5cm]{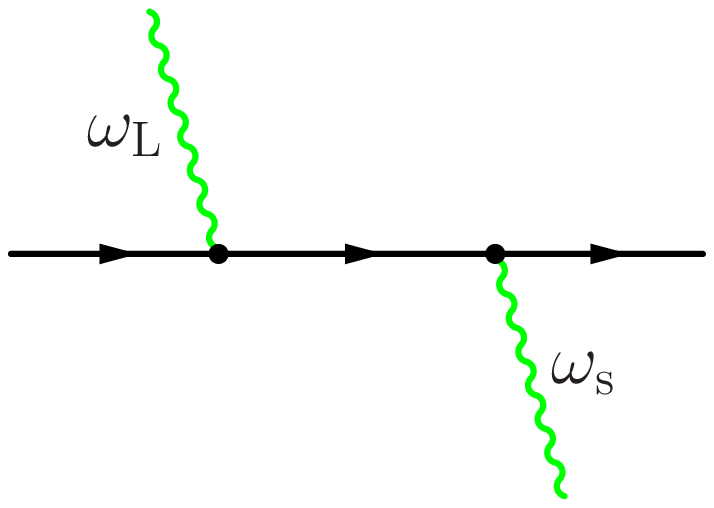}\\
{\sl (a)}
\vskip 0.5cm
\includegraphics[height=2.5cm]{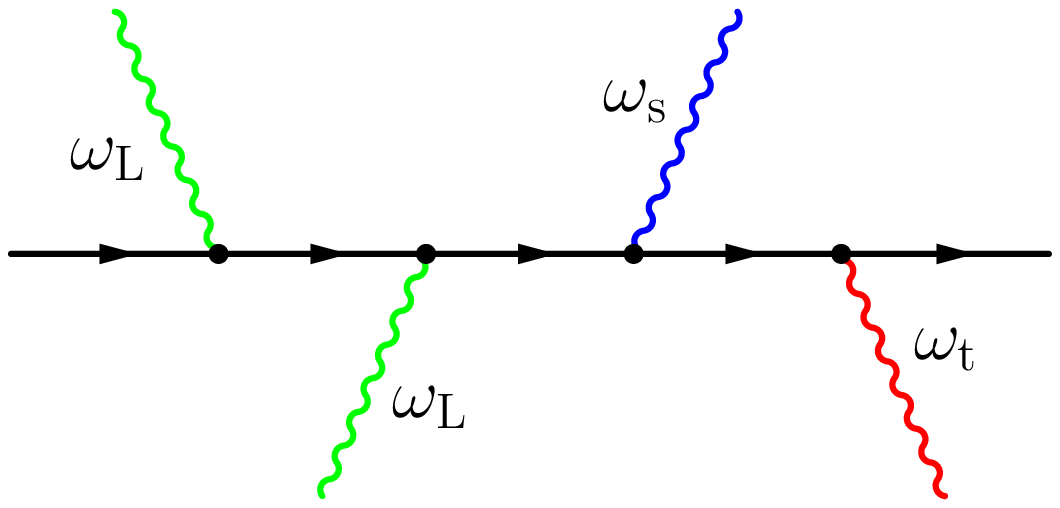}\\
{\sl (b)}\\
\caption{\label{fig1} In an elastic scattering process
[Fig.~{\sl (a)}], the atom absorbs and emits
a single photon, each of frequency $\omega_{\rm L}$.
That is, the atom emits the photon at the 
same frequency as the driving laser frequency.
In an inelastic scattering process [Fig.~{\sl (b)}],
the atom absorbs and emits more than one photon.
Laser frequency $\omega_{\rm L}$. The frequencies
of the scattered photons are $\omega_S$ and 
$\omega_t$.
For many-photon processes, the sum of the emitted photons
equals the sum of the frequencies of the absorbed photons.}
\end{center}
\end{figure}

This entire formalism is not applicable to the case of 
a laser-driven atom in a strong monochromatic (laser) field,
because many-photon processes play a central role
in this regime. The quantum electrodynamic (QED) interaction
would have to be considered in very high orders
of perturbation theory, and this is not feasible 
in practice. One distinguishes between the 
coherently scattered radiation (whose frequency is equal
to the driving frequency) and the incoherently scattered radiation,
which results from the many-photon processes.
For a strong laser field, the ratio of the 
incoherently scattered intensity to the 
coherently scattered intensity
tends to infinity, i.e. the incoherent
part of the atomic fluorescence dominates.

Because it is hopelessly complicated to
try to resum the entire QED series of the many-photon
interactions, one has to invoke a quantum statistical 
treatment which was developed in 
the 1960s and 1970s~\cite{RaSo1962,Mo1969,OlReTa1971,Ba1973};
yet as a considerable simplification, one may 
restrict the Hilbert space of the atom to a few
essential states whose energies are close to resonance.
For instance, we may consider a two-level system
described by the Jaynes--Cummings model~\cite{JaCu1963},
which is
a classic textbook example for a dynamical atom-laser system, well-known in
theoretical quantum optics~\cite{ScZu1997}. 
Due to the driving of the laser field, the atomic
population undergoes Rabi oscillations. The population is driven
periodically from the upper to the lower state and vice versa. The
emission spectrum of this process with a strong driving field is known as
the Mollow spectrum~\cite{Mo1969}; its well-known three-peak
structure may easily
be interpreted in terms of the so-called dressed states,
which are as the eigenstates of the combined system of atom and driving laser
field~\cite{CT1975} in a second-quantized 
formalism. These states diagonalize the atom-field interaction
in all orders of perturbation theory, yet in a truncated
Hilbert space of the atom and within the so-called
rotating-wave approximation. The construction 
of the dressed states also implies
approximations; but these are different from the ones carried 
out in a QED perturbative treatment of the problem and much more
appropriate to the case of a strongly driven atom.
Indeed, the terms left out in carrying out the 
approximations may easily be added later on and lead to 
perturbative corrections to the dressed-state energy levels.
One natural question concerns the coupling of the 
laser-dressed atomic states to the modes of the vacuum
field, i.e.~the Lamb shift of the dressed states.
The appropriate expansion parameters in this context are
the fine-structure constant $\alpha$ and the 
coupling to the atomic nucleus $Z\alpha$.
Furthermore, in a strong field, we may expand
in powers of $\Gamma/\Omega$, where $\Gamma$ is the natural
decay width of the upper level, and $\Omega$ is the 
Rabi frequency, and in $\Omega/\omega_{\rm R}$ and $\Delta/\omega_{\rm R}$,
where $\omega_{\rm R}$ is the atomic resonance 
frequency~\cite{JeEvHaKe2003,JeKe2004aop,EvJeKe2004}.
We hereby assume the Rabi frequency to be large as compared
to the excited-state decay width but small
compared to the atomic transition frequency.
 
We review initially the basic considerations that are relevant to
the description of the Lamb shift of the
laser-dressed states. For a strongly driven
two-level atomic system, one may perform the 
analysis as outlined in Refs.~\cite{JeEvHaKe2003,JeKe2004aop,EvJeKe2004},
using a (two-level) rotating-wave
dressed-state approximation as the starting point.
This leads to a number of intensity- and detuning-dependent 
dynamic corrections to the generalized 
Rabi frequency which determines the Mollow
sidebands. While the bare-state Lamb shift is recovered
in the limit of a vanishing laser intensity,
some of the corrections can only be understood
if one carries out the analysis of the relativistic
and radiative corrections in the dressed-state picture.
Other atomic levels neglected in 
the initial construction of the laser-dressed states,
as well as counter-rotating terms 
and quantum electrodynamic effects
are taken into account perturbatively
and lead to corrections which may be expressed
in terms of the expansion 
parameters $\alpha$, $Z\alpha$, $\Gamma/\Omega$,
$\Omega/\omega_{\rm R}$, and $\Delta/\omega_{\rm R}$.
In this article, we consider a slight generalization of this scheme based on
a laser-driven hydrogenic $1S$--$3P$ 
transition where the $3P$ state may 
decay spontaneously into the metastable $2S$-state.
In the stationary state, a large percentage of the 
atomic population is trapped in the $2S$ state,
and the fluorescence is not very intense. 
However, this should not be an obstacle 
for an experiment provided a suitable intensity-stabilized light
source is available, and we demonstrate here 
that it is possible to generalize the treatment 
of the radiative laser-dressed relativistic and 
radiative corrections 
to multi-level systems.

Experimental work on the Mollow spectrum
crucially depends on the availability of intense
laser light sources because the Mollow side spectrum is only visible
under these conditions (see also Sec. 3.1.2 of~\cite{Mo1981}),
and only if the Doppler shift is essentially eliminated,
which implies the necessity of a collimated atomic beam.
The first experimental confirmations of the 
Mollow theory have been achieved in 
Refs.~\cite{HaWa1973,ScStHe1974,Wa1975,
WuGrEz1975,GrWuEz1975,HaRaScWa1976,GiVe1976,CiGrGaSt1977,GrWuEz1977}.

In this article, we use rationalized Gaussian natural units
with $\hbar = \epsilon_0 = c = 1$. The work is organized as follows:
In Sec.~\ref{basic}, we consider two basic phenomena which are crucial
to an understanding of the interaction of an atom with a quantized
field: {\em (i)} the nonrelativistic part of the self-energy
in Sec.~\ref{nrse}, and {\em (ii)} the dynamic (AC) Stark shift 
of atomic energy levels in a laser field (Sec.~\ref{nrac}). 
Both of these effects are
crucial for the treatment in Sec.~\ref{tr1S3P}, where we 
analyze radiative corrections to the Mollow spectrum for 
hydrogenic $1S$--$3P_j$ transitions ($j = \nicefrac{1}{2},
\nicefrac{3}{2}$) in detail. In Sec.~\ref{sumMollow}, we provide an overview
of all corrections relevant to our analysis. This overview 
is related to the effects discussed in Secs.~\ref{nrse} and~\ref{nrac},
but in addition, we take the opportunity to summarize 
a number of other effects (see also Ref.~\cite{EvJeKe2004}) 
which contribute to the radiative modifications to the Mollow spectrum.
These corrections fall naturally into two groups: the corrections
within the two-level approximation (Sec.~\ref{sumCorr}),
and those beyond this approximation (these result from the multi-level
character of the system due to  
the additional spontaneous decay pathway $3P_j \to 2S \to 1S$, and are
discussed in Sec.~\ref{sumBeyond}). Explicit theoretical predictions
for the $1S$--$3P_j$ transition at specific parameters for the 
Rabi frequency and the detuning are given in Sec.~\ref{explic}.
Finally, conclusions are drawn in Sec.~\ref{conclu}. 

%
%
\section{\label{basic}Basic Atom--Field Interactions}

%
%
\subsection{\label{nrse}Derivation of the Nonrelativistic Part of the Self--Energy}

\begin{figure}[t]
\begin{center}
\includegraphics[width=0.5\linewidth]{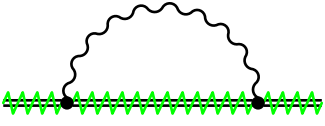}\\
\caption{\label{self-energy}  
The laser-dressed self-energy involves an electron
under the simultaneous influence of a strong driving laser field
(zigzag line) and the binding Coulomb field of the atomic 
nucleus (double line). The formalism used in the analysis
of this term is a generalization of the simple derivation
discussed in Eq.~(\ref{shiftlength}).}
\end{center}
\end{figure}

The nonrelativistic part of the self-energy is the dominant
contribution to the Lamb shift of (bare) atomic 
states due to the bound-state self-energy. 
One might, however, think that the energy shift of the 
electron due to the interaction with the vacuum modes
is actually unobservable as it contributes to its mass.
To resolve this question, it is necessary to consider 
the effects of the binding Coulomb field.  
The self-energy of the bound electron
is then given by the bound-state self-energy shift
minus the corresponding energy shift of a free electron.
The difference is finite and leads to a
small residual effect that shifts the bound-state 
energy levels in comparison to the results
of the Dirac theory.
In a strong laser field, the interaction
with the vacuum modes is modified, because the electron
interacts strongly with the driving laser field
[see Fig.~\ref{self-energy}].

%
%

However, an intuitive understanding can be gained from
a simpler picture.
Consider that in the usual
quantum field theoretic formalism, the
interaction is actually formulated in the
interaction picture, which is why the operators
acquire a time dependence.
There is, however, no reason why one should 
not use the field operators in the time-independent
Schr\"odinger representation. This procedure explicitly 
breaks the covariance,
but one may satisfy oneself that 
Lorentz invariance appears to be broken in bound-state calculations
already via the introduction of the manifestly noncovariant 
Coulomb interaction within the context of a vector potential.
This means that one works in the rest frame of the atomic nucleus,
which is assumed to be infinitely heavy in the non-recoil
limit. In this context, 
one may use stationary as opposed to 
time-dependent perturbation theory,
which simplifies the calculations.

The unperturbed Hamiltonian of the atom is
\begin{equation}
{\mathcal H}_{\rm A} =
\sum_b \, \omega_b \, | b \rangle \, \langle b |\,.
\end{equation}
The normal-ordered Hamiltonian for the electromagnetic field is
given by
\begin{equation}
{\mathcal H}_{\rm F} =
\sum_{\bm{\scriptstyle k}\lambda}
\omega_{\bm{\scriptstyle k}} \,
a^+_{\bm{\scriptstyle k}\lambda} \, a_{\bm{\scriptstyle k}\lambda} \,.
\end{equation}
Therefore, we may assume an unperturbed atom$+$field Hamiltonian
of the form
\begin{equation}
\label{H0}
{\mathcal H}_0 = 
{\mathcal H}_{\rm A} + {\mathcal H}_{\rm F}\,.
\end{equation}
The Schr\"odinger-picture atom-field Hamiltonian in the dipole approximation
and in the length gauge is given by (see~\cite{JeKe2004aop})
\begin{equation}
\label{Hlengthdip}
{\mathcal H}_{\rm AF} \approx 
{\mathcal H}^{\rm (dip)}_{\rm length} = -q \, \bm{x} \cdot \bm{E}\,,
\end{equation}
where $q$ is the physical charge of the 
electron, and the electric-dipole field operator is
\begin{equation}
\bm{E} =
\sum_{\bm{\scriptstyle k}\lambda} 
\frac{1}{\sqrt{V}} \,
\sqrt{\frac{\omega_{\bm{\scriptstyle k}}}{2}} \,
\bm{\epsilon}_\lambda(\bm{\scriptstyle k}) \,
\left[ a_{\bm{\scriptstyle k}\lambda} +
a^+_{\bm{\scriptstyle k}\lambda} \right]\,.
\end{equation}
Here, $a_{\bm{\scriptstyle k}\lambda}$ is the 
discrete-space annihilation operator for a photon
with wave vector $\bm{k}$ and polarization $\lambda \in \{1,2\}$.
It is well known that the dominant contribution to the
Lamb shift (self-energy) is due to
virtual dipole transitions~\cite{EiGrSh2001}. 
The first-order perturbation due to the dipole interaction
Eq.~(\ref{Hlengthdip}) vanishes. The second-order perturbation
(operators in the Schr\"{o}dinger picture) can be written
as
\begin{equation}
\label{shiftlengthdip}
\delta E = \left< a, 0 \left|
{\mathcal H}^{\rm (dip)}_{\rm length} \,
\frac{1}{E_{a,0} - {\mathcal H}_0} \,
{\mathcal H}^{\rm (dip)}_{\rm length} \right|
a, 0 \right>\,,
\end{equation}
where by $| a \rangle$ we denote the atomic reference state
(not to be confused with the creation and annihilation operators!),
and $| b, 0 \rangle$ stands for an atom-field state
with the atom in state $| b \rangle$, and no photons.
Also, we denote by ${\mathcal H}_0$ the unperturbed atom$+$field Hamiltonian
in Eq.~(\ref{H0}). The expression (\ref{shiftlengthdip}) involves a Green 
function which can be written
as a sum over intermediate atom-field states.
A priori, in the intermediate state, we have  the
atom in state $| b \rangle$ and an arbitrary Fock state
of the photon field. However, a nonvanishing contribution
is incurred only from those intermediate
states with one and only one virtual photon.
We denote the wave vector and the polarization state of this single
virtual photon by $\bm{k}\lambda$.

The energy shift Eq.~(\ref{shiftlengthdip}) can be written as
\begin{align}
\label{shiftlength}
E_{\rm SE} &= 
\sum_{b, \bm{\scriptstyle k}\lambda}
\frac{\left< a, 0 \left| {\mathcal H}^{\rm (dip)}_{\rm length} \right| 
b, 1_{\bm{\scriptstyle k}\lambda} \right> \, 
\left< b, 1_{\bm{\scriptstyle k}\lambda} \left| 
{\mathcal H}^{\rm (dip)}_{\rm length} \right| 
a, 0 \right>}{E_a - (E_b + \omega_{\bm{\scriptstyle k}})}
\nonumber\\[2ex]
&= \sum_{b, \bm{\scriptstyle k}\lambda} 
\frac{q^2 \, \omega_{\bm{\scriptstyle k}}}{2\, V} \,
\frac{\left< a \left| 
\bm{\epsilon}_\lambda(\bm{k}) \cdot \bm{x} 
\right| b \right> \, \left< b \left| 
\bm{\epsilon}_\lambda(\bm{k}) \cdot \bm{x} 
\right| a \right>}
{E_a - (E_b + \omega_{\bm{\scriptstyle k}})}
\nonumber\\[2ex]
&= \sum_{b, \bm{\scriptstyle k}} 
\frac{q^2 \, \omega_{\bm{\scriptstyle k}}}{2\, V} \,
\left( \delta^{ij} - \frac{k^i \, k^j}{\bm{k}^2} \right) \,
\frac{\left< a \left| x^i \right| b \right> \, 
\left< b \left| x^j \right| a \right>}
{E_a - (E_b + \omega_{\bm{\scriptstyle k}})}
\nonumber\\[2ex]
&\to q^2 \, \sum_{b} \int \frac{{\rm d}^3 k}{(2\pi)^3} \,
\frac{k}{2} \, \delta^{T,ij} \,
\frac{\left< a \left| x^i \right| b \right> \, 
\left< b \left| x^j \right| a \right>}
{E_a - (E_b + k)}
\nonumber\\[2ex]
&= q^2 \, \sum_{b} \int \frac{{\rm d} k \, k^3}{4 \, \pi^2} \,
\int \frac{{\rm d} \Omega_{\bm{\scriptstyle k}}}{4 \, \pi} \,
\delta^{T,ij} \,
\frac{\left< a \left| x^i \right| b \right> \, 
\left< b \left| x^j \right| a \right>}
{E_a - (E_b + k)}
\nonumber\\[2ex]
&= \frac{2\alpha}{3\,\pi} \, 
\int_0^K {\rm d} k \, k^3 \,
\left< a \left| x^i \, 
\frac{1}{E_a - ({\mathcal H}_{\rm A} + k)} \,
x^i \right| a \right>\,.
\end{align}
Here, 
in going from the third to the fourth line, we applied
the  discrete--continuum transition
\begin{equation}
\label{tocontinuum1}
\sum_{\bm{\scriptstyle k}} \frac{1}{V} \to 
\int \frac{{\rm d}^3 k}{(2\pi)^3} \,,
\end{equation}
which is based on counting the available free photon states in the
normalization volume $V$. 
The transverse $\delta$ function is
\begin{equation}
\delta^{T, ij} = \delta^{ij} - \frac{k^i \, k^j}{\bm{k}^2} \,.
\end{equation}
The integration of the virtual photon energy $k$
from $0$ to $K$ diverges for large $K$; a suitable subtraction
of the first few terms in the asymptotic of the 
integrand for large $k$ leads to a finite result.
The formalism used here is akin
to the first calculation of the Lamb shift by Hans A. Bethe~\cite{Be1947},
and it has recently found a more accurate interpretation in 
terms of methods for the treatment of Lamb shift corrections
inspired by nonrelativistic quantum electrodynamics~\cite{CaLe1986,Pa1993}.
It has the advantage that it can be generalized also 
to other situations.
Indeed, an application of this formalism to the 
laser-dressed states immediately yields the 
dominant self-energy corrections to their
quasi-energies, as outlined in detail in Secs.~4.3 and~4.4 
of Ref.~\cite{JeKe2004aop}. 

%
%
\subsection{\label{nrac}Derivation of the AC Stark Shift}

We start from the second-quantized 
atom-laser Hamiltonian
[see Eq.~(4.10) of~\cite{JeKe2004aop}]:
\begin{equation}
\label{HL}
{\mathcal H}_{\rm L} = -q \, \bm{x} \cdot \bm{E}_{\rm L}\,,
\end{equation}
where the laser-field operator is
\begin{equation}
\label{EL}
\bm{E}_{\rm L} =
\frac{1}{\sqrt{V}} \,
\sqrt{\frac{\omega_{\rm L}}{2}} \,
\bm{\epsilon}_{\rm L} \, \left[ a_{\rm L} + a^+_{\rm L} \right]\,.
\end{equation}
Here, $\bm{\epsilon}_{\rm L} = \hat{z}$ is the laser polarization.
The second-order AC Stark shift is given by
\begin{align}
\label{shiftAC}
E_{\rm AC} =& \sum_b \left\{
\frac{\left< a, n_{\rm L} \left| {\mathcal H}_{\rm L} \right| 
b, n_{\rm L}-1 \right> \, 
\left< b, n_{\rm L}-1 \left| {\mathcal H}_{\rm L} \right| 
a, n_{\rm L} \right>}
{E_a +  n_{\rm L} \,\omega_{\rm L} - (E_b + (n_{\rm L}-1)\,\omega_{\rm L})}
\right.
\nonumber\\[2ex]
& \left. \quad +
\frac{\left< a, n_{\rm L} \left| {\mathcal H}_{\rm L} \right| 
b, n_{\rm L} +1 \right> \, 
\left< b, n_{\rm L} +1 \left| 
{\mathcal H}_{\rm L} \right| 
a, n_{\rm L}  \right>}
{E_a +  n_{\rm L} \,\omega_{\rm L} - (E_b + (n_{\rm L}+1)\,\omega_{\rm L})}
\right\}
\nonumber\\[2ex]
=& \frac{q^2 \, \omega_{\rm L}}{2\, V} \,
\sum_b \left\{
\frac{\left< a \left| z \right| b \right> \, 
\left< b \left| z \right| a \right>}
{E_a - E_b + \omega_{\rm L}} \, n_{\rm L}\right.
\nonumber\\[2ex]
& \left. \quad + \frac{\left< a \left| z \right| b \right> \, 
\left< b \left| z \right| a \right>}
{E_a - E_b - \omega_{\rm L}} \, (n_{\rm L} + 1) \right\}
\nonumber\\[2ex]
=& q^2 \, \frac{n_{\rm L} \, \omega_{\rm L}}{2\, V} \,
\left\{ \left< a \left| z\,\frac{1}{E_a - {\mathcal H}_{\rm A} + \omega_{\rm L}}\,z 
\right| a \right> \right.
\nonumber\\[2ex]
& \quad \left. + \left< a \left| z\,\frac{1}{E_a - {\mathcal H}_{\rm A} - \omega_{\rm L}}\,z
\right| a \right> \right\}
\nonumber\\[2ex]
\to& 2\,\pi\,\alpha\,I \,
\left\{ \left< a \left| z\,\frac{1}{E_a - {\mathcal H}_{\rm A} + \omega_{\rm L}}\,z 
\right| a \right> \right.
\nonumber\\[2ex]
& \quad \left. + \left< a \left| z\,\frac{1}{E_a - {\mathcal H}_{\rm A} - \omega_{\rm L}}\,z
\right| a \right> \right\}
\end{align}
In the transition ``$\to$'' to the continuum limit
$V \to \infty$, which also implies a large 
number of laser photons $n_{\rm L} \to \infty$,
we keep the ratio $n_{\rm L}/V \to$ constant,
as it is this ratio which determines the 
laser intensity. Terms of order $1/V$, which lack
the factor $n_{\rm L}$ in the numerator,
may be neglected in this limit.
In Eq.~(\ref{shiftAC}), 
$I$ denotes the laser intensity, and we remind the reader that 
natural units ($\hbar = c = \epsilon_0 = 1$) are being used.
A generalization of the simple derivation described in 
Eq.~(\ref{shiftAC}) is relevant for the treatment of 
off-resonant corrections (see Sec.~4.6 of~\cite{JeKe2004aop}),
which enter into the spectral decomposition
of the propagator $1/(E_a - {\mathcal H}_{\rm A} - \omega_{\rm L})$.

%
%
\section{\label{tr1S3P}The $1S$--$3P_j$ Transition}

%
%
\subsection{\label{sumMollow}A Summary of the Corrections to the Mollow Spectrum}

\begin{figure}[t]
\begin{center}
\psfrag{Int}[cc][cc][1][0]{Intensity [arb.u.]}
\psfrag{Delta}[cc][cc][1][0]{$\Delta\:[\Gamma]$}
\psfrag{OmegaR}[cc][cc][1][0]{$\Omega_\textrm{R}$}
\includegraphics[height=4cm,width=7cm]{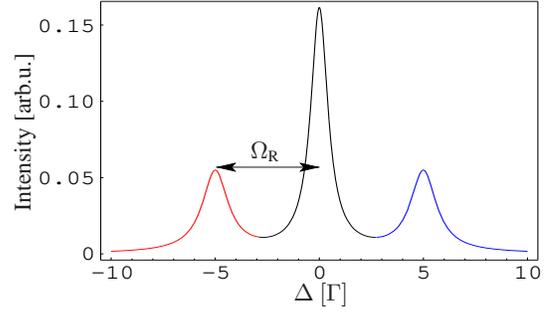}\\
\caption{\label{mollow}  Mollow spectrum.
The incoherently, inelastically scattered intensity is plotted against 
the detuning $\Delta = \omega_{\rm L} - \omega_{\rm R}$,
where $\omega_{\rm L}$ is the laser frequency,
and $\omega_{\rm R}$ is the atomic resonance frequency.
The central elastic peak at zero detuning is not shown
here. The displacement of the red and blue Mollow sideband peaks 
is $\Omega_{\rm R}$ as defined in Eq.~(\ref{OmegaR}).}
\end{center}
\end{figure}

We consider the Mollow spectrum shown in Fig.~\ref{mollow}.
The generalized Rabi frequency, which characterizes the
position of the Mollow sideband peaks relative to the central
peak (see Fig.~\ref{mollow}), is given by
\begin{equation}
\label{OmegaR}
\Omega_{\rm R} = \sqrt{\Omega^2 + \Delta^2} \,.
\end{equation}
Here, the Rabi frequency is
($z$-polarization)
\begin{equation}
\label{defrabi}
\Omega = \langle e | (-q \, z) | g \rangle \,
{\mathcal E}_{\rm L}\,.
\end{equation}
The expression $(\omega_{\rm L}/(2\,V))^{1/2}$ in Eq.~(\ref{EL}) is the
electric laser field ${\mathcal E}_{\rm L}$ per laser photon,
\begin{equation}
\label{fieldperphoton}
{\mathcal E}_{\rm L} = \sqrt{\frac{\omega_{\rm L}}{2 \, V}} \,.
\end{equation}
Its matching with a macroscopic laser field can be done via
\begin{equation}
\label{matchinglaser}
2\,\sqrt{n} \,\, {\mathcal E}_{\rm L} \approx
2\,\sqrt{n+1} \,\, {\mathcal E}_{\rm L}
\to | \bm{{\mathcal E}} | \,,
\end{equation}
where $\bm{{\mathcal E}}$ is the electric field amplitude of the laser
in the convention ${\bm E}_{\rm L}(t) = \bm{{\mathcal E}} \, 
\cos(\omega_{\rm L} \, t)$. 
In the following analysis, we will concentrate on the position
of the sideband peaks in the Mollow spectrum, as it is
a convenient observable. Thus we
consider all corrections as modifications to the  
generalized Rabi frequency $\Omega_{\rm R}$ as defined in Eq.~(\ref{OmegaR}).

%
%

It turns out that all corrections can be interpreted 
as modifications to either the
Rabi frequency $\Omega$ or to the detuning $\Delta$
(see Sec.~3 of~\cite{EvJeKe2004}). These corrections 
will be analyzed here for the $1S$--$3P_j$ transitions
($j = \nicefrac{1}{2}, \nicefrac{3}{2}$) and 
depend on the total angular momentum $j$, i.e.~they 
are different for the two fine-structure components
of the $3P$ state.
Our consideration of this transition is motivated in part
by the recent advent of coherent-wave light sources for the 
hydrogen Lyman-$\alpha$ transition~\cite{EiWaHa2001}.
This means that it is perhaps not unrealistic to consider
the possibility of a coherent-wave light source for the 
$1S$--$3P_j$ transition.
The modifications due to relativistic and radiative 
effects lead to the following modification in
Eq.~(\ref{OmegaR}),
\begin{align}
\label{replDet}
\Delta & \to \Delta - \Delta^{(j)}_{\rm rad}\,,\\[2ex]
\label{replOmega}
\Omega & \to \Omega\, \left (1 + \hat{\Omega}_{\rm rad}^{(j)} \right ) \,.
\end{align}
The $j$-dependent ($j=\nicefrac{1}{2},\nicefrac{3}{2}$) modifications 
are sums of the
various contributions from the considered corrections:
\begin{align}
\label{corrDet}
\Delta^{(j)}_{\rm rad} &=  
L^{(j)}_{\rm bare} + 
{\mathcal B}\, \Omega^2 + 
{\mathcal D}_{\rm R}\, \Omega^2 \,,\\[2ex]
\label{corrOmega}
\hat{\Omega}_{\rm rad}^{(j)} &=
{\mathcal A}_j - {\mathcal C}_j - {\mathcal E}_j  - 
   {\mathcal F} -{\mathcal S}\,,
\end{align}
where all terms are explained in Sec.~\ref{sumCorr} below.
For the moment, we will only remark that the 
corrections to the detuning other than the
bare Lamb shift vanish in the limit of
a negligible laser intensity, i.e. in the 
limit $\Omega \to 0$, as it should be.
Likewise, the relative modification $\hat{\Omega}_{\rm rad}^{(j)}$
of the Rabi frequency has a vanishing influence on the 
Mollow spectrum for small laser intensity because in this
case, the quantity $\Omega$ itself tends to zero. 
The corrected formula for the 
Mollow sideband peaks in the two-level subsystem $\{1S, 3P_j\}$,
\begin{equation}
\label{corrRepl}
\Omega_{\rm R} \to \Omega^{(j)}_{\mathcal C}\,,
\end{equation}
then reads~\cite{EvJeKe2004}
\begin{equation}
\Omega^{(j)}_{\mathcal C} = 
\sqrt{\Omega^2 \cdot \left(1+{\hat{\Omega}_{\rm rad}^{(j)}}\right)^2 +
\left(\Delta - {\Delta^{(j)}_{\rm rad}}\right)^2}\,.
\end{equation}
In view of Eqs.~(\ref{corrDet}) and (\ref{corrOmega}),
the radiative corrections to the detuning lead to a spin-dependent 
dynamic Lamb shift of the Mollow sidebands
\begin{equation}
\label{fullsummation}
\delta \overline{\omega}_\pm^{(j)} = 
\pm \left[ \Omega^{(j)}_{\mathcal C}
\, \left(1 + \hat{\Omega}_{\rm multi} \right)\,
 - \Omega_{\rm R}  \right]\,,
\end{equation}
where the multiplicative modification 
$(1 + \hat{\Omega}_{\rm multi})$ summarizes the effect of 
the additional intermediate atomic levels in a multi-level
configuration (see Sec.~\ref{sumBeyond} below).
This spin-dependent laser-dressed 
Lamb shift depends on two parameters which may be 
dynamically adjusted: the Rabi frequency
and the detuning. From the two-dimensional manifold
spanned by the Rabi frequency
and the detuning, we have picked up one particular parameter 
combination in Sec.~\ref{explic} below.

%
%
\subsection{\label{sumCorr}Corrections Within the Two--Level Approximation}

First, we treat the corrections to the detuning 
listed in Eq.~(\ref{corrDet}).
All corrections summarized here
are discussed in detail in~\cite{EvJeKe2004}.

{\it Bare Lamb shift:} This correction concerns
the term $L^{(j)}_{\rm bare}$ in Eq.~(\ref{corrDet})
and is due to the Lamb shift of the atomic bare states
caused mainly by self-energy corrections due
to interaction of the atom with the surrounding vacuum field.
The result is 
\begin{eqnarray}
\label{Lbarej}
L_{\rm bare}^{(j)} &=& L_{3P_j} - L_{1S} \, .
\end{eqnarray}
For the Lamb shift of the 3P states, we take the data
published in Ref.~\cite{JeSoMo1997}, 
\begin{align}
L_{3P_{\nicefrac{1}{2}}} &= -3473.75(3) \: {\rm kHz},
\\[2ex]
L_{3P_{\nicefrac{3}{2}}} &= 4037.75(3)  \: {\rm kHz}\,.
\end{align}
The $1S$ Lamb shift is given by~\cite{PaJe2003}
\begin{equation}
\label{Lamb-1s}
L_{1S} = 8172811(32)\: {\rm kHz} \,.
\end{equation}

{\it Bloch-Siegert shifts.}
This is the correction term ${\mathcal B}\, \Omega^2$ in Eq.~(\ref{corrDet}),
which is spin-independent to a good approximation~\cite{EvJeKe2004}.
Essentially, the correction is 
caused by counter-rotating atom-field interaction term given by
$\mathcal{H}^{({\rm CR})}_{\rm L} = g_{\rm L} \left (
a^+_{\rm L}|e\rangle\langle g| + a_{\rm L}|g\rangle\langle e|\right )$
(see also Sec.~3 of~\cite{EvJeKe2004} and 
Sec.~4.5 of~\cite{JeKe2004aop}). We only present 
the result here, which reads 
\begin{equation}
{\cal B} = \frac{1}{\omega_{\rm R}} + {\mathcal O}(\Delta/\omega_{\rm L}^2,
\Omega/\omega_{\rm L}^2)\,.
\end{equation}

{\it Off-resonant radiative corrections.}
Here, we are concerned with the term ${\mathcal D}_{\rm R}\, \Omega^2$ in
Eq.~(\ref{corrDet}). The derivation is outlined
in Sec.~4.6 of~\cite{JeKe2004aop} and follows the ideas outlined
in Sec.~\ref{nrac}. The result for this spin-independent term
is (for the notation see Sec.~3 of Ref.~\cite{EvJeKe2004})
\begin{equation}
{\cal D}_{\rm R} = \frac{1}{\omega_{\textrm R}} \: 5.202(3)\,.
\end{equation}
We now turn our attention 
to the various 
relativistic and radiative
corrections to the Rabi frequency listed 
in Eq.~(\ref{corrOmega}).

{\it ${\mathcal E}_j$  - Relativistic corrections to the dipole-moment.}
The evaluation of this spin-dependent
correction to the transition dipole-moment requires
the use of the  relativistic wave functions~\cite{SwDr1991a,SwDr1991b},
see Sec.~3 of~\cite{EvJeKe2004}. The result is
\begin{subequations}
\begin{align}
{\mathcal E}_{\nicefrac{1}{2}} &= (Z\alpha)^2 \,
\left( \frac{17}{24} - \ln 2 + \frac12\, \ln 3\right) \,, \\[2ex]
{\mathcal E}_{\nicefrac{3}{2}} &=  (Z\alpha)^2 \,
\left( \frac{5}{24} - \frac34\,\ln 2 + \frac12\, \ln 3\right) \,.
\end{align}
\end{subequations}

{\it Field-configuration dependent correction.}
This is the ${\mathcal F}$-term in Eq.~(\ref{corrOmega}).
In evaluating this correction, we assume
that the atom is placed in a standing-wave field,
at an  anti-node of the electric field, which implies
also that the magnetic field can be neglected to a very good 
approximation. We are concerned with a field that is
polarized along the $z$-direction,
but whose magnitude is not constant in the propagation direction
The correction may be evaluated using 
the long-wavelength QED Hamiltonian introduced in~\cite{Pa2004}
(see also Sec.~3 of~\cite{EvJeKe2004}).
The result reads
\begin{equation}
{\cal F} = \frac{1}{27}\:(Z\alpha)^2\,.
\end{equation}

{\it Higher-order corrections (in $\Omega$, $\Delta$) to the 
laser-dressed self-energy.}
This correction, which leads to the  ${\mathcal C}_j$--term 
in Eq.~(\ref{corrOmega}), follows from an analysis of 
the Feynman diagram in Fig.~\ref{self-energy}.
The correction has been analyzed in detail 
in~\cite{JeEvHaKe2003}
and in Sec.~4.4 of~\cite{JeKe2004aop}.
The result is 
\begin{equation}
\mathcal{C}_j = 
\alpha (Z\alpha)^2 \: \frac{10}{9\pi}\: 
\left (\ln [(Z\alpha)^{-2}] - 2 \pm 2 \right ) \,,
\end{equation}
where the spin-dependent constant term remains to be evaluated.
We present here an estimate for this term ($-2 \pm 2$);
this estimate is based on
the considerations outlined in Sec.~3 of~\cite{EvJeKe2004}.

{\it Radiative correction to the dipole moment.}
This spin-dependent correction is given by the 
${\mathcal A}_j$-term in Eq.~(\ref{corrOmega}).
The leading logarithmic term, however,
is spin-independent and given by the action of 
the radiative local potential 
\begin{equation}
\label{VLamb}
\Delta V_{\rm lamb}=
\frac{4}{3}\:\alpha (Z\alpha)^2\:\ln[(Z\alpha)^{-2}]\:
\frac{\delta^{(3)}(\bm r)}{m^2}
\end{equation}
on the $1S$ state,
\begin{equation}
\delta d^{(\rm log)} = 
\left<3P \left | z\, \left (\frac{1}{E-H} \right )^{'} 
\Delta V_{{\rm lamb}}({\bm r}) \right |1S\right> \,.
\end{equation}
Note that the radiative potential in Eq.~(\ref{VLamb}) is local,
i.e.~proportional to a Dirac-delta-function, such that the 
corresponding correction to the $3P$ wave function vanishes. 
The result for the spin-independent logarithmic
correction is
\begin{align}
{\cal A}_j = \frac{\alpha\,(Z\alpha)^2}{\pi} \,
 &\left\{ \left[ \frac{55}{27} + 
\frac43\,\ln\left(\frac32\right)\right]\,\ln[(Z\alpha)^{-2}]
\right. \nonumber\\[2ex]
 &  \left. \qquad -5.2 \pm 5.2 \right \}\,,
\end{align}
where we estimate the constant term according to 
the considerations outlined in~\cite{EvJeKe2004}.

{\it ${\mathcal S}$ - Corrections to the secular approximation.}
These are corrections of higher-order  in $(\Gamma / \Omega)$ to the 
expression for the incoherent resonance fluorescence spectrum,
see Eq.~(2.64) of~\cite{JeKe2004aop}.
The result is 
\begin{equation}
{\cal S} = \frac{1}{2}\:\left(\frac{\Gamma}{\Omega}\right)^2
+ {\mathcal O}\left (\frac{\Delta^2\Gamma^2}{\Omega^4}\right)\,.
\end{equation}

%
%
\subsection{\label{sumBeyond}Beyond the Two--Level Approximation:
Corrections Due to the Intermediate $2S$--Level}

%
%
\begin{figure}[t]
\begin{center}
\includegraphics[height=3cm]{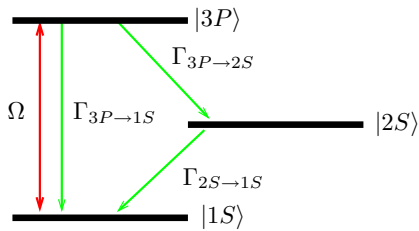}\\
\caption{\label{fig-3level}   
Relevant level scheme of the hydrogen atom. The upper state
$3P$ decays via two channels to the ground state
$1S$; the dominant channel is the direct decay,
while the second channel involves the intermediate $2S$
state. Energy differences are not drawn to scale.}
\end{center}
\end{figure}

Going beyond the two-level approximation, the upper $3P$
state decays via two decay channels as shown in Fig.~\ref{fig-3level}.
The dominant decay chanel is the direct decay to the $1S$ state;
the other channel involves the $2S$ state as intermediate state.
The branching ratios are (see also~\cite{BeSa1957}, p.~266)
\begin{subequations}
\begin{align}
A(3P\to2S) &= 0.2245 \times 10^8 \,\, {\rm s}^{-1}\,,\\[2ex]
A(3P\to1S) & =1.6725 \times 10^8 \,\, {\rm s}^{-1}\,.
\end{align}
\end{subequations}
Qualitatively, the system dynamics is as follows: We assume the
atom to be in the ground state when it enters the interaction
region with the laser field. On the timescale $A(3P\to 1S)^{-1}$,
the atom reaches a quasi-stationary state within the two-level
subspace $(1S,3P)$. At this point, the fluorescence light of
the atom is well described within the above two-level approximation.
Then, on the slower timescale $A(3P\to 2S)^{-1}$, the atom
reaches its true steady state. 
In this stationary state, most of the population is trapped in 
the metastable $2S$ state; but
still the atom emits some fluorescence light whose spectrum deviates 
from the two-level Mollow spectrum due to the third atomic
state as outlined in Ref.~\cite{EvKe2002pra}. 
In particular, the additional decay channel via the intermediate $2S$ 
state induces a relative 
shift of the Mollow sidebands as defined in Eq.~(\ref{fullsummation}) 
given by ($\Omega = 1000\,\Gamma$, $\Delta = 50\,\Gamma$)
\begin{equation}
\label{resMulti}
\hat{\Omega}_{\rm multi} = 6.3 \times 10^{-7}\,.
\end{equation}
In the following, we include this shift as an additional error, as in
an experiment involving a beam of atoms both the two-level and the 
three-level spectra would be observed simultaneously. It should, however,
be noted that the two-level spectrum would clearly 
dominate the experimental outcome,  as the intensity of the three-level 
spectrum is very low due to the trapping of the population 
in the $2S$ state. Therefore, in the final results Eqs.~(\ref{res12},
\ref{res32}), the error induced by the additional atomic state is 
given  separately.

%
%
\subsection{\label{explic}Explicit Values for the 
$1S$--$3P_{1/2}$ and $1S$--$3P_{3/2}$ Transition}

We consider the example $\Omega = 1000\,\Gamma$, $\Delta = 50\,\Gamma$,
but with parameters adjusted for the $1S$--$3P_j$ transition
(cf.~\cite{EvJeKe2004}).
The total correction is [see Table~\ref{tab1} and Eqs.~(\ref{fullsummation}) 
as well as~(\ref{resMulti})]
\begin{equation}
\label{res12}
\delta \overline{\omega}_\pm^{(\nicefrac{1}{2})} = 
1612394(33)(18) \: \textrm{kHz}
\end{equation}
for the $1S$--$3P_{\nicefrac{1}{2}}$ transition
and
\begin{equation}
\label{res32}
\delta \overline{\omega}_\pm^{(\nicefrac{3}{2})} = 
1610305(33)(18)\: \textrm{kHz} 
\end{equation}
for the $1S$--$3P_{\nicefrac{3}{2}}$ transition.
Here, the first bracket denotes the uncertainty
arising from the uncertainties of the individual
correction terms as listed in Tab.~\ref{tab1}, while the second bracket 
corresponds to the uncertainty due to the additional
decay channel via the $2S$ state, see Eqs.~(\ref{fullsummation},
\ref{resMulti}).
Note that the results Eqs.~(\ref{res12},\ref{res32}) deviate 
slightly from the sum of the respective 
corrections in Tab.~\ref{tab1}, as there the
shifts are evaluated individually.

\begin{table}[t]
\begin{ruledtabular}
\begin{center}
\begin{tabular}{|c|.|.|}
\multicolumn{1}{|c|}{\rule[-3mm]{0mm}{8mm}
Shift} &
\multicolumn{1}{c|}{1$S_{\nicefrac{1}{2}} 
\leftrightarrow$ 3$P_{\nicefrac{1}{2}}$ [kHz]} &
\multicolumn{1}{c|}{1$S_{\nicefrac{1}{2}} 
\leftrightarrow$ 3$P_{\nicefrac{3}{2}}$ [kHz]} \\
\hline
\hline
\rule[-3mm]{0mm}{8mm}
$\delta \overline{\omega}^{\rm (Lamb)}_{+,j}$
& 1613618x(11)     & 1611093x(11)  \\
\rule[-3mm]{0mm}{8mm}
$\delta \overline{\omega}^{\rm (BS)}_+$
& -3x.025(1)   & -3x.025(1) \\
\rule[-3mm]{0mm}{8mm}
$\delta \overline{\omega}^{\rm (OR)}_+$
& -62x.91(4)   & -62x.91(4) \\
\hline
\hline
\rule[-3mm]{0mm}{8mm}
$\delta \overline{\omega}^{\rm (R)}_{+,j}$
& -799x.16(4)    & -336x.63(2) \\
\rule[-3mm]{0mm}{8mm}
$\delta \overline{\omega}^{\rm (F)}_{+}$
& -52x.434(3)    & -52x.434(3) \\
\rule[-3mm]{0mm}{8mm}
$\delta \overline{\omega}^{\rm (C)}_{+,j}$
& -29x(6)     &  -29x(6) \\
\rule[-3mm]{0mm}{8mm}
$\delta \overline{\omega}^{\rm (TDM)}_{+,j}$
& 66x(17)    & 66x(17)   \\
\rule[-3mm]{0mm}{8mm}
$\delta \overline{\omega}^{(S)}_{+}$ 
& -13x.29(4)   & -13x.29(4)  \\
\end{tabular}
\end{center}
\caption{\label{tab1}
Summary of all individual energy shifts due to the
various discussed corrections. These are all corrections 
to the Rabi frequency which follow from the analysis outlined
in Sec.~\ref{sumCorr}, i.e.~within the two-level approximation.
For corrections which are beyond this approximation,
see Sec.~\ref{sumBeyond}.}
\end{ruledtabular}
\end{table}

%
%
\section{\label{conclu}Conclusions}

In this article, we have analyzed the relativistic 
and  radiative corrections to the stationary-state quasi-energies
in a hydrogenic multi-level 
$1S$-$3P_{\nicefrac{1}{2}}$-$2S$ configuration.
We demonstrate that it is possible to obtain
theoretical predictions including the corrections
which are beyond the two-level approximation
(see Secs.~\ref{sumCorr} and~\ref{sumBeyond}, respectively).
Explicit predictions for the case $\Omega = 1000\,\Gamma$, 
$\Delta = 50\,\Gamma$ are provided in Eqs.~(\ref{res12})
and~(\ref{res32}).

The modification of the Lamb shift in a dressed environment
(i.e., a strong laser field) is complementary to other ``dressed'' situations
like a cavity. Indeed, the Lamb shift in a cavity
has received considerable attention in the past two decades,
both theoretically as well as 
experimentally~\cite{Ba1970,Ba1987a,Ba1987b,LuRa1985,HeFe1987,SaSuHiHa1992,%
SuEtAl1993,BrEtAl1994,NaLaWa1997}.
Also, one should take the occasion to mention
that laser-dressed states have been used in many 
experiments on the Mollow-related phenomena,
especially dressed states involving Rydberg levels. 
One of the `classic' experimental setups in the field
involves a maser tuned near resonance to a transition between two
Rydberg states, which provides a strong driving field, as well as a microwave
cavity whose eigenmode is also close to the resonance.
There are three frequencies relevant to the problem:
(i) the frequency of the driving field,
(ii) the atomic resonance frequency (corresponding to the
transition between Rydberg states),
and (iii) the eigenmode of the cavity.
It is well known that spontaneous emission can be enhanced if the
cavity eigenmode frequency is equal to a Mollow sideband~\cite{ZhLeLe1988}.
The modifications of the cavity-induced spontaneous
emission are well described by the optical master
equation~\cite{AgLaWa1993}, and good agreement between theory and experiment
is obtained~\cite{AgLaWa1993,LaWa1993}.
For example, in Ref.~\cite{LaAgWa1996}, the authors
describe an experiment in which a two-photon transition between
dressed states is observed in a driven microwave cavity.
The atomic beam consists of ${}^{85}{\rm Rb}$ atoms
excited to the $53 {}^2 {\rm P}_{3/2}$ state, and a maser is tuned
near, but not on resonance to the transition
to the $53 {}^2 {\rm S}_{1/2}$ state.
When the detuning $\Delta$ is adjusted so that
$\Delta = \pm \Omega_{\rm R}/2$, the otherwise dominant one-photon
decay is suppressed, and the atom decays via emission of two
photons into the cavity mode, under simultaneous absorption of
a photon from the laser mode. The third photon is necessary in order
to ensure angular momentum conservation in the electric dipole
transition.
Consequently, a further natural ground for an extension of the 
ideas outlined here would be to consider the laser-dressed Lamb shift
in the additionally modified environment of a cavity.

In general, our approach is concerned with 
atomic physics; yet the analysis
requires the techniques of quantum field theory.
Approximate answers can be obtained for realistic
situations by simply applying the results of the asymptotic-state
formalism~\cite{GMLo1951,Su1957} 
to the quantities that are of relevance to the dynamical process.
However, as demonstrated here and in~\cite{JeEvHaKe2003,JeKe2004aop,%
EvJeKe2004}, there exist some small residual dynamic effects which can
only be understood if one uses a picture in which
the dynamics is treated in full.
An experimental investigation of the laser-dressed Lamb shift,
which depends on the laser intensity and the detuning,
would require an accurate measurement of a (stabilized) laser
intensity, which is problematic; other experimental issues
include the Doppler effect, which must be controlled 
by using a well-collimated atomic beam~\cite{HaRaScWa1976}.
Yet, the intriguing consequences of the theoretical predictions may 
warrant this effort. Recent progress in heterodyne 
measurements of the resonance fluorescence of single 
ions~\cite{HoBaLaWa1997,HoBaEiHeWa1997} may indicate that
ultimately, achieving high accuracy in the measurement 
of the fluorescence spectra will rather depend on technical
questions than on fundamental theoretical limitations.


\begin{acknowledgments}
Financial support is acknowledged by the German Science
Foundation for the authors, and by the German National 
Academic Foundation for J.E..
The authors acknowledge helpful discussions with M. Haas
regarding the evaluation of the matrix elements for the
off-resonant corrections.
\end{acknowledgments}


\begin{thebibliography}{51}
\expandafter\ifx\csname natexlab\endcsname\relax\def\natexlab#1{#1}\fi
\expandafter\ifx\csname bibnamefont\endcsname\relax
  \def\bibnamefont#1{#1}\fi
\expandafter\ifx\csname bibfnamefont\endcsname\relax
  \def\bibfnamefont#1{#1}\fi
\expandafter\ifx\csname citenamefont\endcsname\relax
  \def\citenamefont#1{#1}\fi
\expandafter\ifx\csname url\endcsname\relax
  \def\url#1{\texttt{#1}}\fi
\expandafter\ifx\csname urlprefix\endcsname\relax\def\urlprefix{URL }\fi
\providecommand{\bibinfo}[2]{#2}
\providecommand{\eprint}[2][]{\url{#2}}

\bibitem[{\citenamefont{Kramers and Heisenberg}(1925)}]{KrHe1925}
\bibinfo{author}{\bibfnamefont{W.}~\bibnamefont{Kramers}} \bibnamefont{and}
  \bibinfo{author}{\bibfnamefont{W.~H.} \bibnamefont{Heisenberg}},
  \bibinfo{journal}{Z. Phys.} \textbf{\bibinfo{volume}{31}},
  \bibinfo{pages}{681} (\bibinfo{year}{1925}).

\bibitem[{\citenamefont{Weisskopf}(1931)}]{We1931}
\bibinfo{author}{\bibfnamefont{V.}~\bibnamefont{Weisskopf}},
  \bibinfo{journal}{Ann. Phys. (Leipzig)} \textbf{\bibinfo{volume}{9}},
  \bibinfo{pages}{23} (\bibinfo{year}{1931}).

\bibitem[{\citenamefont{Gell-Mann and Low}(1951)}]{GMLo1951}
\bibinfo{author}{\bibfnamefont{M.}~\bibnamefont{Gell-Mann}} \bibnamefont{and}
  \bibinfo{author}{\bibfnamefont{F.}~\bibnamefont{Low}},
  \bibinfo{journal}{Phys. Rev.} \textbf{\bibinfo{volume}{84}},
  \bibinfo{pages}{350} (\bibinfo{year}{1951}).

\bibitem[{\citenamefont{Sucher}(1957)}]{Su1957}
\bibinfo{author}{\bibfnamefont{J.}~\bibnamefont{Sucher}},
  \bibinfo{journal}{Phys. Rev.} \textbf{\bibinfo{volume}{107}},
  \bibinfo{pages}{1448} (\bibinfo{year}{1957}).

\bibitem[{\citenamefont{Rautian and Sobel'man}(1962)}]{RaSo1962}
\bibinfo{author}{\bibfnamefont{S.~G.} \bibnamefont{Rautian}} \bibnamefont{and}
  \bibinfo{author}{\bibfnamefont{I.~I.} \bibnamefont{Sobel'man}},
  \bibinfo{journal}{JETP} \textbf{\bibinfo{volume}{14}}, \bibinfo{pages}{328}
  (\bibinfo{year}{1962}).

\bibitem[{\citenamefont{Mollow}(1969)}]{Mo1969}
\bibinfo{author}{\bibfnamefont{B.~R.} \bibnamefont{Mollow}},
  \bibinfo{journal}{Phys. Rev.} \textbf{\bibinfo{volume}{188}},
  \bibinfo{pages}{1969} (\bibinfo{year}{1969}).

\bibitem[{\citenamefont{Oliver et~al.}(1971)\citenamefont{Oliver, Ressayre, and
  Tallet}}]{OlReTa1971}
\bibinfo{author}{\bibfnamefont{G.}~\bibnamefont{Oliver}},
  \bibinfo{author}{\bibfnamefont{E.}~\bibnamefont{Ressayre}}, \bibnamefont{and}
  \bibinfo{author}{\bibfnamefont{A.}~\bibnamefont{Tallet}},
  \bibinfo{journal}{Lett. Nuovo Cimento} \textbf{\bibinfo{volume}{2}},
  \bibinfo{pages}{777} (\bibinfo{year}{1971}).

\bibitem[{\citenamefont{Baklanov}(1973)}]{Ba1973}
\bibinfo{author}{\bibfnamefont{E.~V.} \bibnamefont{Baklanov}},
  \bibinfo{journal}{Zh. \'{E}ksp. Teor. Fiz.} \textbf{\bibinfo{volume}{65}},
  \bibinfo{pages}{2203} (\bibinfo{year}{1973}), \bibinfo{note}{[JETP {\bf 38},
  1100 (1974)]}.

\bibitem[{\citenamefont{Jaynes and Cummings}(1963)}]{JaCu1963}
\bibinfo{author}{\bibfnamefont{E.~T.} \bibnamefont{Jaynes}} \bibnamefont{and}
  \bibinfo{author}{\bibfnamefont{F.~W.} \bibnamefont{Cummings}},
  \bibinfo{journal}{Proc. IEEE} \textbf{\bibinfo{volume}{51}},
  \bibinfo{pages}{89} (\bibinfo{year}{1963}).

\bibitem[{\citenamefont{Scully and Zubairy}(1997)}]{ScZu1997}
\bibinfo{author}{\bibfnamefont{M.~O.} \bibnamefont{Scully}} \bibnamefont{and}
  \bibinfo{author}{\bibfnamefont{M.~S.} \bibnamefont{Zubairy}},
  \emph{\bibinfo{title}{Quantum Optics}} (\bibinfo{publisher}{Cambridge
  University Press}, \bibinfo{address}{Cambridge}, \bibinfo{year}{1997}).

\bibitem[{\citenamefont{Cohen-Tannoudji}(1975)}]{CT1975}
\bibinfo{author}{\bibfnamefont{C.}~\bibnamefont{Cohen-Tannoudji}},
  \emph{\bibinfo{title}{Atoms in Strong Resonant Fields}}
  (\bibinfo{publisher}{North-Holland}, \bibinfo{address}{Amsterdam},
  \bibinfo{year}{1975}), pp. \bibinfo{pages}{4--104}.

\bibitem[{\citenamefont{Jentschura et~al.}(2003)\citenamefont{Jentschura,
  Evers, Haas, and Keitel}}]{JeEvHaKe2003}
\bibinfo{author}{\bibfnamefont{U.~D.} \bibnamefont{Jentschura}},
  \bibinfo{author}{\bibfnamefont{J.}~\bibnamefont{Evers}},
  \bibinfo{author}{\bibfnamefont{M.}~\bibnamefont{Haas}}, \bibnamefont{and}
  \bibinfo{author}{\bibfnamefont{C.~H.} \bibnamefont{Keitel}},
  \bibinfo{journal}{Phys. Rev. Lett.} \textbf{\bibinfo{volume}{91}},
  \bibinfo{pages}{253601} (\bibinfo{year}{2003}).

\bibitem[{\citenamefont{Jentschura and Keitel}(2004)}]{JeKe2004aop}
\bibinfo{author}{\bibfnamefont{U.~D.} \bibnamefont{Jentschura}}
  \bibnamefont{and} \bibinfo{author}{\bibfnamefont{C.~H.}
  \bibnamefont{Keitel}}, \bibinfo{journal}{Ann. Phys. (N.Y.)}
  \textbf{\bibinfo{volume}{310}}, \bibinfo{pages}{1} (\bibinfo{year}{2004}).


\bibitem[{\citenamefont{Evers et~al.}()\citenamefont{Evers, Jentschura, and
  Keitel}}]{EvJeKe2004}
\bibinfo{author}{\bibfnamefont{J.}~\bibnamefont{Evers}},
  \bibinfo{author}{\bibfnamefont{U.~D.} \bibnamefont{Jentschura}},
  \bibnamefont{and} \bibinfo{author}{\bibfnamefont{C.~H.}
  \bibnamefont{Keitel}},\bibinfo{journal}{Phys. Rev. A}
  \textbf{\bibinfo{volume}{70}}, \bibinfo{pages}{062111} (\bibinfo{year}{2004}).

\bibitem[{\citenamefont{Mollow}(1981)}]{Mo1981}
\bibinfo{author}{\bibfnamefont{B.~R.} \bibnamefont{Mollow}}, in
  \emph{\bibinfo{booktitle}{Progress in Optics XIX}}, edited by
  \bibinfo{editor}{\bibfnamefont{E.}~\bibnamefont{Wolf}}
  (\bibinfo{publisher}{North-Holland}, \bibinfo{address}{Amsterdam},
  \bibinfo{year}{1981}), pp. \bibinfo{pages}{2--43}.

\bibitem[{\citenamefont{Hartig and Walther}(1973)}]{HaWa1973}
\bibinfo{author}{\bibfnamefont{W.}~\bibnamefont{Hartig}} \bibnamefont{and}
  \bibinfo{author}{\bibfnamefont{H.}~\bibnamefont{Walther}},
  \bibinfo{journal}{Ann. Phys. (N.Y.)} \textbf{\bibinfo{volume}{1}},
  \bibinfo{pages}{171} (\bibinfo{year}{1973}).

\bibitem[{\citenamefont{Schuda et~al.}(1974)\citenamefont{Schuda, Strout, and
  Hercher}}]{ScStHe1974}
\bibinfo{author}{\bibfnamefont{F.}~\bibnamefont{Schuda}},
  \bibinfo{author}{\bibfnamefont{C.~R.} \bibnamefont{Strout}},
  \bibnamefont{and} \bibinfo{author}{\bibfnamefont{M.}~\bibnamefont{Hercher}},
  \bibinfo{journal}{J. Phys. B} \textbf{\bibinfo{volume}{7}},
  \bibinfo{pages}{L198} (\bibinfo{year}{1974}).

\bibitem[{\citenamefont{Walther}(1975)}]{Wa1975}
\bibinfo{author}{\bibfnamefont{H.}~\bibnamefont{Walther}}, in
  \emph{\bibinfo{booktitle}{Proc. Second Laser Spectroscopy Conf.}}, edited by
  \bibinfo{editor}{\bibfnamefont{S.}~\bibnamefont{Haroche}},
  \bibinfo{editor}{\bibfnamefont{J.~C.} \bibnamefont{Pebay-Peroula}},
  \bibinfo{editor}{\bibfnamefont{T.~W.} \bibnamefont{H\"{a}nsch}},
  \bibnamefont{and} \bibinfo{editor}{\bibfnamefont{S.~H.} \bibnamefont{Harris}}
  (\bibinfo{publisher}{Springer}, \bibinfo{address}{Heidelberg},
  \bibinfo{year}{1975}), pp. \bibinfo{pages}{358--369}.

\bibitem[{\citenamefont{Wu et~al.}(1975)\citenamefont{Wu, Grove, and
  Ezekiel}}]{WuGrEz1975}
\bibinfo{author}{\bibfnamefont{F.~Y.} \bibnamefont{Wu}},
  \bibinfo{author}{\bibfnamefont{R.~E.} \bibnamefont{Grove}}, \bibnamefont{and}
  \bibinfo{author}{\bibfnamefont{S.}~\bibnamefont{Ezekiel}},
  \bibinfo{journal}{Phys. Rev. Lett.} \textbf{\bibinfo{volume}{35}},
  \bibinfo{pages}{1426} (\bibinfo{year}{1975}).

\bibitem[{\citenamefont{Grove et~al.}(1975)\citenamefont{Grove, Wu, and
  Ezekiel}}]{GrWuEz1975}
\bibinfo{author}{\bibfnamefont{R.}~\bibnamefont{Grove}},
  \bibinfo{author}{\bibfnamefont{F.~Y.} \bibnamefont{Wu}}, \bibnamefont{and}
  \bibinfo{author}{\bibfnamefont{S.}~\bibnamefont{Ezekiel}},
  \bibinfo{journal}{Phys. Rev. A} \textbf{\bibinfo{volume}{15}},
  \bibinfo{pages}{227} (\bibinfo{year}{1975}).

\bibitem[{\citenamefont{Hartig et~al.}(1976)\citenamefont{Hartig, Rasmussen,
  Schieder, and Walther}}]{HaRaScWa1976}
\bibinfo{author}{\bibfnamefont{W.}~\bibnamefont{Hartig}},
  \bibinfo{author}{\bibfnamefont{W.}~\bibnamefont{Rasmussen}},
  \bibinfo{author}{\bibfnamefont{R.}~\bibnamefont{Schieder}}, \bibnamefont{and}
  \bibinfo{author}{\bibfnamefont{H.}~\bibnamefont{Walther}},
  \bibinfo{journal}{Z. Phys. A} \textbf{\bibinfo{volume}{278}},
  \bibinfo{pages}{205} (\bibinfo{year}{1976}).

\bibitem[{\citenamefont{Gibbs and Venkatesan}(1976)}]{GiVe1976}
\bibinfo{author}{\bibfnamefont{H.~M.} \bibnamefont{Gibbs}} \bibnamefont{and}
  \bibinfo{author}{\bibfnamefont{T.~N.~C.} \bibnamefont{Venkatesan}},
  \bibinfo{journal}{Opt. Commun.} \textbf{\bibinfo{volume}{17}},
  \bibinfo{pages}{87} (\bibinfo{year}{1976}).

\bibitem[{\citenamefont{Citron et~al.}(1977)\citenamefont{Citron, Gray, Gabel,
  and Stroud}}]{CiGrGaSt1977}
\bibinfo{author}{\bibfnamefont{M.~L.} \bibnamefont{Citron}},
  \bibinfo{author}{\bibfnamefont{H.~R.} \bibnamefont{Gray}},
  \bibinfo{author}{\bibfnamefont{C.~W.} \bibnamefont{Gabel}}, \bibnamefont{and}
  \bibinfo{author}{\bibfnamefont{C.~R.} \bibnamefont{Stroud}},
  \bibinfo{journal}{Phys. Rev. A} \textbf{\bibinfo{volume}{16}},
  \bibinfo{pages}{1507} (\bibinfo{year}{1977}).

\bibitem[{\citenamefont{Grove et~al.}(1977)\citenamefont{Grove, Wu, and
  Ezekiel}}]{GrWuEz1977}
\bibinfo{author}{\bibfnamefont{R.~E.} \bibnamefont{Grove}},
  \bibinfo{author}{\bibfnamefont{F.~Y.} \bibnamefont{Wu}}, \bibnamefont{and}
  \bibinfo{author}{\bibfnamefont{S.}~\bibnamefont{Ezekiel}},
  \bibinfo{journal}{Phys. Rev. A} \textbf{\bibinfo{volume}{15}},
  \bibinfo{pages}{227} (\bibinfo{year}{1977}).

\bibitem[{\citenamefont{Eides et~al.}(2001)\citenamefont{Eides, Grotch, and
  Shelyuto}}]{EiGrSh2001}
\bibinfo{author}{\bibfnamefont{M.~I.} \bibnamefont{Eides}},
  \bibinfo{author}{\bibfnamefont{H.}~\bibnamefont{Grotch}}, \bibnamefont{and}
  \bibinfo{author}{\bibfnamefont{V.~A.} \bibnamefont{Shelyuto}},
  \bibinfo{journal}{Phys. Rep.} \textbf{\bibinfo{volume}{342}},
  \bibinfo{pages}{63} (\bibinfo{year}{2001}).

\bibitem[{\citenamefont{Bethe}(1947)}]{Be1947}
\bibinfo{author}{\bibfnamefont{H.~A.} \bibnamefont{Bethe}},
  \bibinfo{journal}{Phys. Rev.} \textbf{\bibinfo{volume}{72}},
  \bibinfo{pages}{339} (\bibinfo{year}{1947}).

\bibitem[{\citenamefont{Caswell and Lepage}(1986)}]{CaLe1986}
\bibinfo{author}{\bibfnamefont{W.~E.} \bibnamefont{Caswell}} \bibnamefont{and}
  \bibinfo{author}{\bibfnamefont{G.~P.} \bibnamefont{Lepage}},
  \bibinfo{journal}{Phys. Lett. B} \textbf{\bibinfo{volume}{167}},
  \bibinfo{pages}{437} (\bibinfo{year}{1986}).

\bibitem[{\citenamefont{Pachucki}(1993)}]{Pa1993}
\bibinfo{author}{\bibfnamefont{K.}~\bibnamefont{Pachucki}},
  \bibinfo{journal}{Ann. Phys. (N.Y.)} \textbf{\bibinfo{volume}{226}},
  \bibinfo{pages}{1} (\bibinfo{year}{1993}).

\bibitem[{\citenamefont{Eikema et~al.}(2001)\citenamefont{Eikema, Walz, and
  H\"{a}nsch}}]{EiWaHa2001}
\bibinfo{author}{\bibfnamefont{K.~S.~E.} \bibnamefont{Eikema}},
  \bibinfo{author}{\bibfnamefont{J.}~\bibnamefont{Walz}}, \bibnamefont{and}
  \bibinfo{author}{\bibfnamefont{T.~W.} \bibnamefont{H\"{a}nsch}},
  \bibinfo{journal}{Phys. Rev. Lett.} \textbf{\bibinfo{volume}{86}},
  \bibinfo{pages}{5679} (\bibinfo{year}{2001}).

\bibitem[{\citenamefont{Jentschura et~al.}(1997)\citenamefont{Jentschura, Soff,
  and Mohr}}]{JeSoMo1997}
\bibinfo{author}{\bibfnamefont{U.~D.} \bibnamefont{Jentschura}},
  \bibinfo{author}{\bibfnamefont{G.}~\bibnamefont{Soff}}, \bibnamefont{and}
  \bibinfo{author}{\bibfnamefont{P.~J.} \bibnamefont{Mohr}},
  \bibinfo{journal}{Phys. Rev. A} \textbf{\bibinfo{volume}{56}},
  \bibinfo{pages}{1739} (\bibinfo{year}{1997}).

\bibitem[{\citenamefont{Pachucki and Jentschura}(2003)}]{PaJe2003}
\bibinfo{author}{\bibfnamefont{K.}~\bibnamefont{Pachucki}} \bibnamefont{and}
  \bibinfo{author}{\bibfnamefont{U.~D.} \bibnamefont{Jentschura}},
  \bibinfo{journal}{Phys. Rev. Lett.} \textbf{\bibinfo{volume}{91}},
  \bibinfo{pages}{113005} (\bibinfo{year}{2003}).

\bibitem[{\citenamefont{Swainson and Drake}(1991{\natexlab{a}})}]{SwDr1991a}
\bibinfo{author}{\bibfnamefont{R.~A.} \bibnamefont{Swainson}} \bibnamefont{and}
  \bibinfo{author}{\bibfnamefont{G.~W.~F.} \bibnamefont{Drake}},
  \bibinfo{journal}{J. Phys. A} \textbf{\bibinfo{volume}{24}},
  \bibinfo{pages}{79} (\bibinfo{year}{1991}{\natexlab{a}}).

\bibitem[{\citenamefont{Swainson and Drake}(1991{\natexlab{b}})}]{SwDr1991b}
\bibinfo{author}{\bibfnamefont{R.~A.} \bibnamefont{Swainson}} \bibnamefont{and}
  \bibinfo{author}{\bibfnamefont{G.~W.~F.} \bibnamefont{Drake}},
  \bibinfo{journal}{J. Phys. A} \textbf{\bibinfo{volume}{24}},
  \bibinfo{pages}{95} (\bibinfo{year}{1991}{\natexlab{b}}).

\bibitem[{\citenamefont{Pachucki}(2004)}]{Pa2004}
\bibinfo{author}{\bibfnamefont{K.}~\bibnamefont{Pachucki}},
  \bibinfo{journal}{Phys. Rev. A} \textbf{\bibinfo{volume}{69}},
  \bibinfo{pages}{052502} (\bibinfo{year}{2004}).

\bibitem[{\citenamefont{Bethe and Salpeter}(1957)}]{BeSa1957}
\bibinfo{author}{\bibfnamefont{H.~A.} \bibnamefont{Bethe}} \bibnamefont{and}
  \bibinfo{author}{\bibfnamefont{E.~E.} \bibnamefont{Salpeter}},
  \emph{\bibinfo{title}{Quantum Mechanics of One- and Two-Electron Atoms}}
  (\bibinfo{publisher}{Springer}, \bibinfo{address}{Berlin},
  \bibinfo{year}{1957}).

\bibitem[{\citenamefont{Evers and Keitel}(2002)}]{EvKe2002pra}
\bibinfo{author}{\bibfnamefont{J.}~\bibnamefont{Evers}} \bibnamefont{and}
  \bibinfo{author}{\bibfnamefont{C.~H.} \bibnamefont{Keitel}},
  \bibinfo{journal}{Phys. Rev. A} \textbf{\bibinfo{volume}{65}},
  \bibinfo{pages}{033813} (\bibinfo{year}{2002}).

\bibitem[{\citenamefont{Barton}(1970)}]{Ba1970}
\bibinfo{author}{\bibfnamefont{G.}~\bibnamefont{Barton}},
  \bibinfo{journal}{Proc. Roy. Soc. London A} \textbf{\bibinfo{volume}{320}},
  \bibinfo{pages}{251} (\bibinfo{year}{1970}).

\bibitem[{\citenamefont{Barton}(1987{\natexlab{a}})}]{Ba1987a}
\bibinfo{author}{\bibfnamefont{G.}~\bibnamefont{Barton}},
  \bibinfo{journal}{Proc. Roy. Soc. London A} \textbf{\bibinfo{volume}{410}},
  \bibinfo{pages}{141} (\bibinfo{year}{1987}{\natexlab{a}}).

\bibitem[{\citenamefont{Barton}(1987{\natexlab{b}})}]{Ba1987b}
\bibinfo{author}{\bibfnamefont{G.}~\bibnamefont{Barton}},
  \bibinfo{journal}{Proc. Roy. Soc. London A} \textbf{\bibinfo{volume}{410}},
  \bibinfo{pages}{175} (\bibinfo{year}{1987}{\natexlab{b}}).

\bibitem[{\citenamefont{L\"{u}tken and Ravndal}(1985)}]{LuRa1985}
\bibinfo{author}{\bibfnamefont{C.~A.} \bibnamefont{L\"{u}tken}}
  \bibnamefont{and} \bibinfo{author}{\bibfnamefont{F.}~\bibnamefont{Ravndal}},
  \bibinfo{journal}{Phys. Rev. A} \textbf{\bibinfo{volume}{31}},
  \bibinfo{pages}{2082} (\bibinfo{year}{1985}).

\bibitem[{\citenamefont{Heinzen and Feld}(1987)}]{HeFe1987}
\bibinfo{author}{\bibfnamefont{D.~J.} \bibnamefont{Heinzen}} \bibnamefont{and}
  \bibinfo{author}{\bibfnamefont{M.~S.} \bibnamefont{Feld}},
  \bibinfo{journal}{Phys. Rev. Lett.} \textbf{\bibinfo{volume}{59}},
  \bibinfo{pages}{2623} (\bibinfo{year}{1987}).

\bibitem[{\citenamefont{Sandoghdar et~al.}(1992)\citenamefont{Sandoghdar,
  Sukenik, Hinds, and Haroche}}]{SaSuHiHa1992}
\bibinfo{author}{\bibfnamefont{V.}~\bibnamefont{Sandoghdar}},
  \bibinfo{author}{\bibfnamefont{C.~I.} \bibnamefont{Sukenik}},
  \bibinfo{author}{\bibfnamefont{E.~A.} \bibnamefont{Hinds}}, \bibnamefont{and}
  \bibinfo{author}{\bibfnamefont{S.}~\bibnamefont{Haroche}},
  \bibinfo{journal}{Phys. Rev. Lett.} \textbf{\bibinfo{volume}{68}},
  \bibinfo{pages}{3432} (\bibinfo{year}{1992}).

\bibitem[{\citenamefont{Sukenik et~al.}(1993)\citenamefont{Sukenik, Boshier,
  Cho, Sandoghdar, and Hinds}}]{SuEtAl1993}
\bibinfo{author}{\bibfnamefont{C.~I.} \bibnamefont{Sukenik}},
  \bibinfo{author}{\bibfnamefont{M.~G.} \bibnamefont{Boshier}},
  \bibinfo{author}{\bibfnamefont{D.}~\bibnamefont{Cho}},
  \bibinfo{author}{\bibfnamefont{V.}~\bibnamefont{Sandoghdar}},
  \bibnamefont{and} \bibinfo{author}{\bibfnamefont{E.~A.} \bibnamefont{Hinds}},
  \bibinfo{journal}{Phys. Rev. Lett.} \textbf{\bibinfo{volume}{70}},
  \bibinfo{pages}{560} (\bibinfo{year}{1993}).

\bibitem[{\citenamefont{Brune et~al.}(1994)\citenamefont{Brune, Nussenzveig,
  Schmidt-Kaler, Bernardot, Maali, Raimond, and Haroche}}]{BrEtAl1994}
\bibinfo{author}{\bibfnamefont{M.}~\bibnamefont{Brune}},
  \bibinfo{author}{\bibfnamefont{P.}~\bibnamefont{Nussenzveig}},
  \bibinfo{author}{\bibfnamefont{F.}~\bibnamefont{Schmidt-Kaler}},
  \bibinfo{author}{\bibfnamefont{F.}~\bibnamefont{Bernardot}},
  \bibinfo{author}{\bibfnamefont{A.}~\bibnamefont{Maali}},
  \bibinfo{author}{\bibfnamefont{J.~M.} \bibnamefont{Raimond}},
  \bibnamefont{and} \bibinfo{author}{\bibfnamefont{S.}~\bibnamefont{Haroche}},
  \bibinfo{journal}{Phys. Rev. Lett.} \textbf{\bibinfo{volume}{72}},
  \bibinfo{pages}{3339} (\bibinfo{year}{1994}).

\bibitem[{\citenamefont{Nakajima et~al.}(1997)\citenamefont{Nakajima,
  Lambropoulos, and Walther}}]{NaLaWa1997}
\bibinfo{author}{\bibfnamefont{T.}~\bibnamefont{Nakajima}},
  \bibinfo{author}{\bibfnamefont{P.}~\bibnamefont{Lambropoulos}},
  \bibnamefont{and} \bibinfo{author}{\bibfnamefont{H.}~\bibnamefont{Walther}},
  \bibinfo{journal}{Phys. Rev. A} \textbf{\bibinfo{volume}{56}},
  \bibinfo{pages}{5100} (\bibinfo{year}{1997}).

\bibitem[{\citenamefont{Zhu et~al.}(1988)\citenamefont{Zhu, Lezama, Lewenstein,
  and Mossberg}}]{ZhLeLe1988}
\bibinfo{author}{\bibfnamefont{Y.}~\bibnamefont{Zhu}},
  \bibinfo{author}{\bibfnamefont{A.}~\bibnamefont{Lezama}},
  \bibinfo{author}{\bibfnamefont{M.}~\bibnamefont{Lewenstein}},
  \bibnamefont{and} \bibinfo{author}{\bibfnamefont{T.~W.}
  \bibnamefont{Mossberg}}, \bibinfo{journal}{Phys. Rev. Lett.}
  \textbf{\bibinfo{volume}{61}}, \bibinfo{pages}{1946} (\bibinfo{year}{1988}).

\bibitem[{\citenamefont{Agarwal et~al.}(1993)\citenamefont{Agarwal, Lange, and
  Walther}}]{AgLaWa1993}
\bibinfo{author}{\bibfnamefont{G.~S.} \bibnamefont{Agarwal}},
  \bibinfo{author}{\bibfnamefont{W.}~\bibnamefont{Lange}}, \bibnamefont{and}
  \bibinfo{author}{\bibfnamefont{H.}~\bibnamefont{Walther}},
  \bibinfo{journal}{Phys. Rev. A} \textbf{\bibinfo{volume}{48}},
  \bibinfo{pages}{4555} (\bibinfo{year}{1993}).

\bibitem[{\citenamefont{Lange and Walther}(1993)}]{LaWa1993}
\bibinfo{author}{\bibfnamefont{W.}~\bibnamefont{Lange}} \bibnamefont{and}
  \bibinfo{author}{\bibfnamefont{H.}~\bibnamefont{Walther}},
  \bibinfo{journal}{Phys. Rev. A} \textbf{\bibinfo{volume}{48}},
  \bibinfo{pages}{4551} (\bibinfo{year}{1993}).

\bibitem[{\citenamefont{Lange et~al.}(1996)\citenamefont{Lange, Agarwal, and
  Walther}}]{LaAgWa1996}
\bibinfo{author}{\bibfnamefont{W.}~\bibnamefont{Lange}},
  \bibinfo{author}{\bibfnamefont{G.~S.} \bibnamefont{Agarwal}},
  \bibnamefont{and} \bibinfo{author}{\bibfnamefont{H.}~\bibnamefont{Walther}},
  \bibinfo{journal}{Phys. Rev. Lett.} \textbf{\bibinfo{volume}{76}},
  \bibinfo{pages}{3293} (\bibinfo{year}{1996}).

\bibitem[{\citenamefont{H\"offges
  et~al.}(1997{\natexlab{a}})\citenamefont{H\"offges, Baldauf, Lange, and
  Walther}}]{HoBaLaWa1997}
\bibinfo{author}{\bibfnamefont{J.~T.} \bibnamefont{H\"offges}},
  \bibinfo{author}{\bibfnamefont{H.~W.} \bibnamefont{Baldauf}},
  \bibinfo{author}{\bibfnamefont{W.}~\bibnamefont{Lange}}, \bibnamefont{and}
  \bibinfo{author}{\bibfnamefont{H.}~\bibnamefont{Walther}},
  \bibinfo{journal}{J. Mod. Opt.} \textbf{\bibinfo{volume}{44}},
  \bibinfo{pages}{1999} (\bibinfo{year}{1997}{\natexlab{a}}).

\bibitem[{\citenamefont{H\"offges
  et~al.}(1997{\natexlab{b}})\citenamefont{H\"offges, Baldauf, Eichler,
  Helmfrid, and Walther}}]{HoBaEiHeWa1997}
\bibinfo{author}{\bibfnamefont{J.~T.} \bibnamefont{H\"offges}},
  \bibinfo{author}{\bibfnamefont{H.~W.} \bibnamefont{Baldauf}},
  \bibinfo{author}{\bibfnamefont{T.}~\bibnamefont{Eichler}},
  \bibinfo{author}{\bibfnamefont{S.~R.} \bibnamefont{Helmfrid}},
  \bibnamefont{and} \bibinfo{author}{\bibfnamefont{H.}~\bibnamefont{Walther}},
  \bibinfo{journal}{Opt. Commun.} \textbf{\bibinfo{volume}{133}},
  \bibinfo{pages}{170} (\bibinfo{year}{1997}{\natexlab{b}}).

\end{thebibliography}
\end{document}